\documentstyle[aaspp4]{article}

\def \etal{et~al.\/}

\begin{document}

\title{Analytic Inversion of Emission Lines of Arbitrary Optical Depth
for the Structure of Supernova Ejecta}

\author{R.\ Ignace\altaffilmark{1,}\altaffilmark{2} \& 
        M.\ A.\ Hendry\altaffilmark{3,}\altaffilmark{4} }
 
\altaffiltext{1}{
        Email:  ri@astro.physics.uiowa.edu }
 
\altaffiltext{2}{
	203 Van Allen Hall, 
	Department of Physics and Astronomy,
	University of Iowa, 
	Iowa City, IA 52242 USA }

\altaffiltext{3}{
        Email:  martin@astro.gla.ac.uk } 

\altaffiltext{4}{
        Department of Physics and Astronomy,
        Kelvin Building,
        University of Glasgow,
        Glasgow G12 8QQ,
        Scotland UK }

\begin{abstract}

We derive a method for inverting emission line profiles formed in
supernova ejecta.  The derivation assumes spherical symmetry and
homologous expansion (i.e., $v(r) \propto r$), is analytic, and even
takes account of occultation by a pseudo-photosphere.  Previous
inversion methods have been developed which are restricted to optically
thin lines, but the particular case of homologous expansion permits an
analytic result for lines of {\it arbitrary} optical depth.  In fact,
we show that the quantity that is generically retrieved is the run of
line intensity $I_\lambda$ with radius in the ejecta.  This result is
quite general, and so could be applied to resonance lines,
recombination lines, etc.  As a specific example, we show how to derive
the run of (Sobolev) optical depth $\tau_\lambda$ with radius in the
case of a pure resonance scattering emission line.


\end{abstract}

\section{INTRODUCTION}

For many years analysis of the spectra of various types of supernovae
has been a highly active area of research, with relevance to studies of
nucleosynthesis and galactic chemical evolution, stellar winds and the
interstellar medium, and the extragalactic distance scale (c.f.,
Chevalier 1990; Eastman, Schmidt, \& Kirshner 1996; Fillipenko 1997;
Branch 1998). Fast numerical codes (c.f., Fisher 1999) now exist to
``forward'' model synthetic spectra -- which can then be matched to
observed spectra -- from a parametrized treatment of the temperature,
velocity, density and composition structure of supernova ejecta (c.f.,
Millard \etal\ 1999). On the other hand, there has been comparatively
little study of the {\em inverse\/} approach, to infer the radial
dependence of intrinsic physical characteristics of the ejecta directly
from spectral line profiles.  Fransson \& Chevalier (1989) and Fransson
(1994) investigate the use of optically thin line profile shapes as a
diagnostic of the emissivity with radius for supernova ejecta in
homologous expansion (i.e., $v(r) \propto r$).  However, we realized
that an analytic inversion for emission lines of arbitrary optical
depth could be derived.

As described in this paper, our inversion relies on the fact that the
emission from a geometrically thin shell with an expansion speed that
is linear in radius will produce a rectangular or ``flat-topped''
emission profile (such a profile shape is more commonly associated with
optically thin lines formed in constant expansion shells).  After
undertaking this study, we discovered a comment by Kuhi (1973) who
discussed flat-topped emission lines observed in the spectra of
Wolf-Rayet stars.  Appearing in a ``Note added in proof'', Kuhi credits
C.~Magnan for pointing out that an optically thick spherical shell with
$v \propto r$ can also produce a flat-topped profile.  Unfortunately,
no reference to the remark is given, nor have we been able to locate
any similar conclusions in other published works.

The paper is organized as follows.  Section~2 presents a derivation of
the inversion technique, beginning first with a discussion of Sobolev
theory in Section~2.1, then a derivation of the basic inversion in 2.2,
followed by a discussion of occultation effects in 2.3, and ending with
an illustrative application to a pure resonance scattering line in
2.4.  A discussion of the method and caveats to practical applications
are given in Section~3.

\section{THE LINE INVERSION}

We consider ejecta from a supernova as existing in a spherical shell in
homologous expansion with inner radius $r_{\rm min}$ and outer radius
$r_{\rm max}$.  There may or may not be a pseudo-photosphere existing
in this shell (a pseudo-photosphere will exist at early times following
the supernova, but fade toward late times).  Allowing for a
pseudo-photosphere, we define $r_{\rm ph}$ as its radius.  Given these
assumptions, we show how to derive information about the radial
structure of the ejecta, in terms of density, ionization structure,
heating input, and so on, or some combination thereof from the
inversion of an observed emission line profile.  This inversion is
applicable to lines of {\it arbitrary optical depth}.  The formalism
follows arguments developed by Brown \etal\ (1997) and Ignace
\etal\ (1998) for the inversion of optically thin emission lines formed
in stellar winds.  Whereas those authors were seeking to derive the
wind speed distribution for a known stellar mass loss rate, here we can
assume a speed distribution $v(r) \propto r$ and thus solve for other
properties in the ejecta shell.

\subsection{Sobolev Theory}

We employ standard Sobolev theory (Sobolev 1960; Mihalas 1978; Lamers
\& Cassinelli 1999).  A small parcel of gas at radius $r$ from the
center of the supernova (taken to be the origin)
and expanding radially with $v(r)$ will produce
line emission with a Doppler shift given by $\Delta \lambda =
\lambda_0\,\frac{v(r)}{c}\,\cos \theta$, where $\theta$ is the
spherical polar angle between the sightline (taken to be the $z-$axis)
and the line from the origin to the gas parcel.  For convenience, we
use $\mu=\cos\theta$ hereafter and also introduce the line-of-sight
velocity shift $v_{\rm z} = v(r)\,\mu$.

Sobolev theory provides the following expressions for the radiative
transfer in a spherical expanding envelope.  The flux is given by

\begin{equation}
F_\lambda (\Delta \lambda) = \frac{1}{D^2}\, \int_{\Delta \lambda}\,
	2\pi\,I_\lambda\,p\,dp,
\end{equation}

\noindent where $p$ is the impact parameter orthogonal to the $z-$axis,
$D$ the distance, $I_\lambda$ the monochromatic line intensity, and the
integration proceeds over the isovelocity zone specified by $\Delta
\lambda$ a constant.  We assume a velocity law of the form $v(r) =
v_{\rm max} \, (r/r_{\rm max})$, where $v_{\rm max}$ is the maximum
speed obtained by the shell located at largest radius $r_{\rm max}$.
The locus of points of constant Doppler shift $\Delta \lambda =
\lambda_0\,v_{\rm z}/c$ is given by $v_{\rm z} = v_{\rm
max}\,\mu\,(r/r_{\rm max}) = v_{\rm max}\,(z/r_{\rm max})$.  This means
that isovelocity zones are planes oriented orthogonally to the
observer's line-of-sight.  Because the ejecta resides in a shell of
finite extent, these zones are disks or annuli that cut through the
shell.  (Annuli occur when the disk cuts through the shell for
$z<r_{\rm min}$ or $z<r_{\rm ph}$, thereby intersecting either the
inner evacuated region of the shell or the pseudo-photosphere.)

The intensity from Sobolev theory is $I_\lambda = S_\lambda \, \left(1
- e^{-\tau_\lambda}\right)$, where the source function $S_\lambda$
and the Sobolev optical depth $\tau_\lambda$ are evaluated at the point
of consideration.  The optical depth is given by

\begin{equation}
\tau_\lambda (p) = \frac{\kappa_\lambda(r)\,\rho(r)\,
	\nu_0}{| dv_{\rm z}/dz |},
\end{equation}

\noindent where $\kappa_\lambda(r)$ is the line opacity, $\rho(r)$ the
mass density in the shell, and $dv_{\rm z}/dz$ is the line-of-sight
velocity gradient.  For a spherical flow, this velocity gradient is

\begin{equation}
\frac{dv_{\rm z}}{dz} = \mu^2\,\frac{dv}{dr}+ (1-\mu^2)\,\frac{v}{r}.
	\label{eq:dvdz}
\end{equation}

\noindent Using the preceding expressions from Sobolev theory,
we next formulate the line inversion technique.

\subsection{The Basic Inversion Result}

It is important to recall the well-known result that a constant
expansion shell producing optically thin line emission will have a
flat-topped profile shape.  Such a line will have emission spread
between velocity shifts $v_{\rm z} = + v(r)$ and $v_{\rm z} = - v(r)$.
For an optically thin line formed in a shell with a range of radial
speeds, one can construct the line profile as a superposition of
flat-tops, each arising from a sub-shell expanding at constant speed.
The different flat-top emission contributions have width in wavelength
as determined by the speed of the sub-shell and emission height
determined by its volume and emissivity.  The inversion methods derived
by Brown \etal\ (1997) and Ignace \etal\ (1998) are based on this kind
of construction.  Differentiating the emission profile, their technique
is to use the observed Doppler shift to trace out the physical velocity
in the flow, so that the profile slope essentially yields information
on the run of density (or more generally the emissivity) with radius.
They show that for a given emissivity, one can invert the emission
profile to yield $v(r)$.

For arbitrary optical depths, the contribution from each shell is not
generally flat-topped, thus one cannot usually employ this inversion.
However, the case of linear or homologous expansion is a special case,
with each shell producing a flat-topped emission contribution for
arbitrary optical depth (as we next show), and hence an inversion is
still possible.  Consider equation~(\ref{eq:dvdz}) for the velocity
gradient. In the case of linear expansion, the gradient reduces to
$dv_{\rm z}/dz = \mu^2\,(v_{\rm max}/r_{\rm max}) + (1-\mu^2)\,(v_{\rm
max}/r_{\rm max}) = v_{\rm max}/r_{\rm max}$.  This is constant,
meaning that the Sobolev optical depth $\tau_\lambda = \tau_\lambda(r)$
is a function of radius only.  Scattering or production of photons in a
flow with a velocity gradient will normally be anisotropic, but the
case of homologous expansion is special, yielding an optical depth that
is isotropic.  In the language of Sobolev theory, this means that the 
escape probability is likewise isotropic.  Since the source
function is a function of radius, the intensity is consequently a
function of radius too, with $I_\lambda(r) = S_\lambda(r)\,[ 1-
e^{\tau_\lambda(r)}]$.

Now consider a geometrically thin spherical shell undergoing homologous
expansion.  The flux of line emission will be $F_\lambda (\Delta
\lambda) = (2\pi/D^2)\, I_\lambda(r)\,p\,dp$.  The isovelocity zones
are rings oriented with constant $z$.  However, we also have the
geometric relation that $r^2 = p^2 +z^2$, and since $z={\rm{constant}}$ at
fixed $\Delta \lambda$, we find that $r\,dr = p\,dp$.  Thus the flux
becomes $F_\lambda (\Delta \lambda) = (2\pi/D^2)\,
I_\lambda(r)\,r\,dr$.  The result is that the flux appearing at any
$\Delta \lambda$ is a function of radius only, hence the flux at all
$\Delta \lambda$ in the line profile is the same --
a flat-topped emission profile, even though we have nowhere required
the line to be optically thin.

Extending this to a shell of arbitrary extent with a linear velocity
law as specified, we have the total flux at Doppler shift $\Delta
\lambda$ being

\begin{equation}
F_\lambda (\Delta \lambda) = \frac{2\pi}{D^2}\,\int_{r(v)}^{r_{\rm max}}\, 
        I_\lambda(r)\,r\,dr.
	\label{eq:flux}
\end{equation}

\noindent Making a change of coordinate from $r$ to $v$, with
$dr = dv/(dv/dr)$, we obtain

\begin{equation}
F_\lambda (\Delta \lambda) = \frac{2\pi}{D^2}\,\int_{v}^{v_{\rm max}}\, 
        I_\lambda(r(v))\,r(v)\,\frac{dv}{dv/dr}.
\end{equation}
 
\noindent Differentiating with respect to $\Delta \lambda$ and noting
that $d\lambda = d(\Delta \lambda)$ and $dv/dr = v_{\rm max}/r_{\rm
max} \equiv t^{-1}_0$, we find that

\begin{equation}
\left( \frac{dF_\lambda}{d\lambda}\right)_{\Delta \lambda}
	= -\frac{2\pi\,t_0}{D^2}\,r\,I_\lambda(r)\, \left(
	\frac{dv}{d\lambda}\right)_{\Delta \lambda}
	= -\frac{2\pi\,c\,t_0}{\lambda_0\,D^2}\,r\,I_\lambda(r),
\end{equation}

\noindent where $t_0$ is the age of the supernova ejecta shell at the
time of the line measurement, and the derivatives are evaluated at the
point in the line profile where $\Delta \lambda = \lambda_0\,v(r)/c$.
The minus sign signifies that the profile gradient $dF_\lambda/d\lambda
<0$ (i.e., emission increases from the wings toward line center).  The
lefthand side is the data, and so we can directly solve for the
intensity on the righthand side.

\subsection{Occultation Effects}

We have so far ignored the pseudo-photosphere, which can influence our
results by way of line absorption and geometric occultation.  As is
standard for P Cygni profiles, the column of material intervening
between the photosphere and the observer will scatter or absorb
photospheric light.  Since this column of gas is moving toward the
observer, the absorption will appear as blueshifted from line center.
As a consequence, the blueshifted part of the profile consists of the
emission part superposed on the absorption trough.  The net result is a
distorted profile shape, which may be in absorption or emission.  The
resolution to this difficulty is simply to apply our inversion to the
redshifted part of the profile only.

However, geometric occultation modifies the shape of the redshifted
profile by virtue of blocking out line emission from the receding
column of gas on the far side of the photosphere.  If we can
approximate the photosphere as a spherical ball of infinite optical
depth, we can correct for the occultation as follows (e.g., see Ignace
\etal\ 1998).  Even with occultation, the emission from a shell still
produces a flat-top profile, except that the emission at greatest
redshift is no longer observed.  The maximum redshift for emission from
a shell of radius $r$ no longer extends to $v(r)$ but rather
$\tilde{v}(r) = v(r)\,\sqrt{1-r_{\rm ph}^2/r^2}$, purely from
considerations of geometry.  So $\tilde{v}(r)$ is the redshift of gas
where the shell of radius $r$ intersects the occultation tube that
extends behind the pseudo-photosphere.  As expected, a shell very near
the photosphere will be almost entirely occulted, for which $\tilde{v}
\approx 0$, whereas a shell of radius $r\gg r_{\rm ph}$ will suffer
little occultation and thus $\tilde{v} \approx v$.

Substituting $\tilde{v}$ for $v$ and applying the inversion to
equation~(\ref{eq:flux}) for the flux of line emission in the
redshifted profile, we obtain

\begin{equation}
\left(\frac{dF_\lambda}{d\lambda}\right)_{\Delta \lambda = \lambda_0\,\tilde{v}/c}
	= -\frac{2\pi\,c\,t_0}{\lambda_0\,D^2}\,r\,I_\lambda(r)\,
	\sqrt{1-\frac{r_{\rm ph}^2}{r^2}}, 
\end{equation}

\noindent where we have substituted for $d\tilde{v}/dr =
(\tilde{v}/r)\, (1 -r_{\rm ph}^2/r^2)^{-1}$.  To simplify the
expression, we define a normalized emission profile gradient as
$\Phi(\Delta \lambda) = (\lambda_0/F_{\lambda,c}) \,
(dF_\lambda/d\lambda)$, where the continuum flux is $F_{\lambda,c} =
\pi\,I_{\rm ph}\,r_{\rm ph}^2/D^2$.  Solving for the intensity, we get

\begin{equation}
\frac{I_\lambda(r)}{I_{\rm ph}} = \frac{-\Phi(\Delta \lambda)\,r_{\rm ph}^2}
	{2c\,t_0\,r}\,\left(1-\frac{r_{\rm ph}^2}{r^2}\right)^{-1/2}\, 
	\label{eq:occ}
\end{equation}

\noindent where the line profile gradient is evaluated
at $\Delta \lambda = \lambda_0\,\tilde{v}(r)/c $.

\subsection{Application to a Pure Resonance Scattering Line}

As an example, we consider homologous expansion and a pure resonance
scattering line.  A major simplification occurring in this case is that
the source function reduces to the form $S(r) = W(r)\, I_{\rm ph}$, for
the dilution factor $W(r) = 0.5\,\left(1-\sqrt{1-r_{\rm
ph}^2/r^2}\right)$ (e.g., Mihalas 1978).  In this case the intensity
becomes $I_\lambda(r) = W(r)\, I_{\rm ph}\,(1-e^{-\tau_\lambda})$, and
our inversion equation~(\ref{eq:occ}) can be used to obtain the line
optical depth, viz

\begin{equation}
\tau_\lambda(r) = - \ln \left\{ 1-\frac{r_{\rm ph}^2}
	 {2c\,t_0\,r}\cdot\frac{[-\Phi(r)]\,}{[1-2W(r)]\,W(r)}\right\}.
	\label{eq:resline}
\end{equation}

This technique is an extremely powerful diagnostic, {\it if} both
spherical symmetry and homologous expansion may be assumed.  For
example, the optical depth depends on opacity and density associated
with the ion and transition being considered.  Depending on what is
known or can be assumed, the inversion of the line profile can provide
constraints for the mass spectrum of the ejecta, ionization
and/or excitation of the ion, heating from radioactive decay, or some
combination of these properties (e.g., see Fransson \& Chevalier 1989
for similar considerations in the context of their analysis for
optically thin lines).  It is important to stress that the inversion is
model independent (proviso the assumptions) and so can be used to test
current models for supernova explosions.  The applicability of the
inversion requires monotonicity in the velocity field, but the derived
optical depth structure itself need not be monotonic.  Any such
non-monotonicity might for example be a signature of changes in
the ionization of the ion producing the line emission, or perhaps
non-monotonicity in the mass spectrum.

\section{DISCUSSION}

We have shown that emission line profiles of arbitrary optical depth
formed in supernova ejecta shells can be analytically inverted to yield
information about the shell structure.  The method even allows for
geometric occultation by a pseudo-photosphere.  However, there are
a few important checks and caveats that should be noted.

\begin{itemize}

\item[a)] It is key that the method be applied only to emission
lines that are ``pristine'' (i.e., not blended with other lines),
otherwise the profile shape will be affected, perhaps severely,
and the inversion will yield a result, but it won't be physically
meaningful.

\item[b)] We argued that one should apply the inversion to the redshifted
profile only, because absorption on the blueshifted half of the line
would render the inversion technique invalid.  However, having
obtained, for example, the optical depth structure from the redshifted
profile, one could use that information to forward model the blueshifted
profile.  This would provide a check for self-consistency.  Failure
to recover the observed blueshifted profile would imply that either
the shell is not spherically symmetric or the expansion is not homologous.

\item[c)] The emission profile should show a flat-top at line core.  This
flat-top would correspond to the emission from the inner most
observable shell, either at the radius $r_{\rm min}$ with radial speed
$v_{\rm min}$, or effectively just outside the pseudo-photosphere
$r_{\rm ph}$ with radial speed $v_{\rm ph}$.  However, it may be that
the central flat-top could be narrow and difficult to resolve
spectroscopically, or even perhaps that thermal or turbulent broadening
causes the it to be somewhat more ``rounded'' than flat.  In this case
the inversion will artificially tend to yield a result for radii
interior to either $r_{\rm min}$ or $r_{\rm ph}$, as the case may be.
A resolution of this problem is to make use of the {\it total} line
emission of the redshifted wing.  One would take the results from the
inversion, integrate in radius, and compare with the observed line
emission.  The radius at which the integration matches the total
observed line emission should be $r_{\rm min}$ or $r_{\rm ph}$, as
appropriate, and thus where the results from the inversion should be
truncated.

\item[d)] Finally, we wish to point out that at certain times, especially
early times, electron scattering might significantly alter the line
profile shape (e.g., Fransson \& Chevalier 1989).  Electron scattering
will lead to a broadening of the emission line, but in such a way as to
conserve total line flux.  This will alter the profile shape, and hence
bias application of our inversion.  Essentially, the expectation is for
the profile to be broadened and the slope to be diminished.  The
recovered intensity will then be reduced relative to what it would be
in the absence of electron scattering.

\end{itemize}

In conclusion, our inversion technique is a fairly general method for
obtaining structure information about supernova ejecta shells; however,
care must be exercised in the line selection and consideration given to
the time $t_0$ since the explosion, since this latter quantity is
relevant for the prominence of a pseudo-photosphere and the significance
of electron scattering broadening.  Under favorable conditions, our
method should provide important constraints for hydrodynamic and
radiative transfer models of supernova explosions, and possibly in novae
if the required conditions are met.

\acknowledgements

The authors gratefully acknowledge helpful discussions with Profs.\
John Brown, Joe Cassinelli, Jay Gallagher, Ken Gayley, and Bob Mutel.
We also thank an anonymous referee.  MAH acknowledges support for this
research from a PPARC grant.  RI acknowledges funding support from
an NSF grant (AST-9986915).

\end{document}